\documentclass{article}

\usepackage{hyperref}

\usepackage{graphicx}

\usepackage{amsmath}

\newcommand{\bfr}{\mathbf{r}}
\newcommand{\bfu}{\mathbf{u}}

\begin{document}

\title{%
  Rapid convergence for simulations that project from particles onto a
  fixed mesh}

\author{Daniel Duque \& Pep Espa\~nol \\
  Model Basin Research Group (CEHINAV). \\
  ETSI Navales, Universidad Polit\'ecnica de Madrid \\
  Madrid, Spain \\
  \href{mailto:daniel.duque@upm.es}{daniel.duque@upm.es} \\
  \& \\
  Dpto. de F{\'\i}sica Fundamental \\
  Universidad Nacional de Educaci\'on a Distancia \\
  Madrid, Spain
}

\date{\today}

\maketitle

\begin{abstract}
  The advantage of particle Lagrangian methods in computational fluid
  dynamics is that advection is accurately modeled. However, the
  calculation of space derivatives is complicated. On the other hand,
  Eulerian formulations benefit from a fixed mesh, but feature
  non-linear advection terms. It seems natural to combine these two,
  using particles to advect the information and a fixed mesh to
  calculate space derivatives. This idea goes back to Particle-in-Cell
  methods, and is here considered within the context of the finite
  element (FE) method for the fixed mesh, and the particle-FEM (pFEM)
  for the particles.  A ``projection'' method is required to transfer
  field values from particles to mesh and vice versa --- in this work
  simple interpolation is used. Our results confirm that projection
  errors, especially from particles to mesh, cause a slow convergence
  of the method if standard, linear, FEs, are employed --- but also
  that the convergence rate is restored to values comparable to
  Lagrangian simulations if quadratically consistent shape functions
  for the particles are used. The same applies to computational
  resources, making projection a viable alternative to Lagrangian
  simulations. The procedure is validated on Zalesak's disk problem,
  the Taylor-Green vortex sheet flow, and the Rayleigh-Taylor
  instability.
\end{abstract}

\maketitle

\section{Introduction}

In Lagrangian particle methods of computational fluid dynamics,
advection is modelled by moving the particles along the velocity
field. The equations of motion are free of explicit advective terms,
which would in general require a special treatment
\cite{versteeg_2007}. On the other hand, the initial particle setup
can become highly distorted, which can lead to a degradation of the
quality of the space discretization.  An Eulerian method, on the other
hand, benefits from the setup being defined once, at the beginning of
the calculation, usually on a prescribed mesh. The equations of
motion, on the other hand, contain non-linear advection terms, and the
resulting computational problems are likewise not so well posed.  It
therefore seems natural to combine the two approaches, using particles
to advect the field information with time, and keeping the static mesh
to compute space derivatives.  This idea is not new, as it goes back
to Particle-in-Cell methods \cite{PIC, PIC2}.  In the context of SPH
similar methods have been presented, beginning at least with the
remeshing technique of \cite{chaniotis_2002}, and particle splitting
\cite{Feldman_Bonet_2007}. To be precise, these SPH works do not use
an Eulerian mesh, but they do concern the problem of interpolating the
field values at points different from the position of the particles,
which is the crucial problem addressed below.  There has been much
recent effort trying to bring together Lagrangian and Eulerian
simulations \cite{marrone_2015,quinlan_2014}.
%

\begin{figure}
  \centering
  \includegraphics[width=0.4\textwidth]{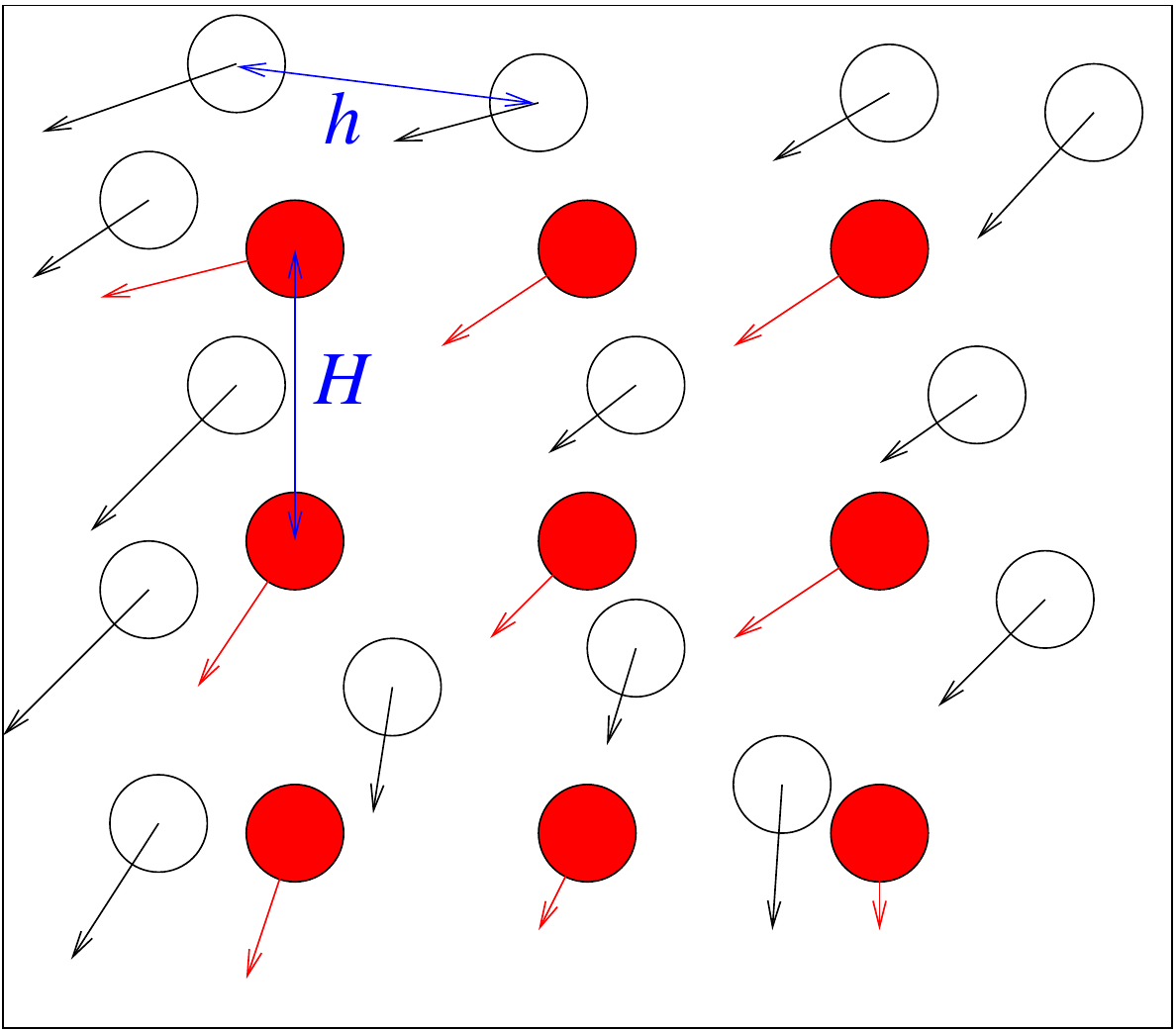}
  \caption{Main idea behind particle-in-cell methods. Fields (here, a
    vector field such as the velocity) are projected to and from the
    nodes of a fixed mesh (in red), with spacing $H$, onto particle
    nodes (in black), with mean spacing $h$.\label{fig:idea}}
\end{figure}

In Fig. \ref{fig:idea} we plot a sketch of this idea. The information
of a field may be transferred from the nodes of a mesh onto the
particles, or vice versa. A vector field (such as the velocity or the
pressure gradient) is depicted, but the same idea applies to scalar
fields, such as the pressure or the density. We will employ $H$ to
refer to the fixed mesh spacing, and $h$ for the mean interparticle
distance.

In \cite{Idelsohn_2015} this idea is explored, in the context of the
finite element method (FEM) for the fixed mesh, and the particle FEM
(pFEM) for the particles. We here follow these ideas, seeking to
improve the errors associated with the interpolation (also referred to
as projection) from particles to mesh and back.  This source of error
was identified, among others, as the one responsible for slow
convergence of results. This finding is confirmed here, and we present
a method that uses quadratically consistent interpolation from the
particles to the mesh that leads to improved convergence.  We want to
remark that both that in those works and here, a simple interpolation
procedure is used for projection. Other, more refined procedures may
be studied, such as Galerkin projection, a technique developed to
reduce the dimensionality of a problem in an optimal way
\cite{stanton2013non,Rapun2017}.

This  article is  structured  as follows.   The necessary  theoretical
background  is  laid  out  in Section  \ref{sec:theory},  featuring  a
general introduction to  integration errors and a  discussion of shape
functions. Results  are presented in \ref{sec:results},  for Zalesak's
disk  (Section  \ref{sec:zalesak}),   the  Taylor-Green  vortex  sheet
(Section \ref{sec:TG}),  and the Rayleigh-Taylor  instability (Section
\ref{sec:RT}).    We  end   up  with   some  conclusions   in  Section
\ref{sec:conclusion}.   The  technical  details  concerning  numerical
methods and the  procedure to obtain quadratic functions  are given in
the Appendices.

\section{Theory}
\label{sec:theory}

\subsection{Integration errors}
\label{sec:errors}

There are several sources of error associated to the spatial and
temporal discretization of the continuous equations
\cite{Idelsohn_2015}. Let us write down a general equation for a
field $A$ as
\begin{align}
  \frac{d A}{d t} &= Q \\
  \label{eq:position_integration}
  \frac{d \bfr }{d t} &= \mathbf{u} 
\end{align}
where the time derivative is the convective derivative and $Q$ is a
source term, which may include space derivatives of $A$. For a
second-order integrator (such as the midpoint method) we will have, at
each time iteration, an error appearing as
\[
Ạ_{n+1} = Ạ_{n} + (\bar{Q} \pm \epsilon_t ) \Delta t ,
\]
where $\Delta t$ is the time-step, which we assume is fixed, $\bar{Q}$
is some mean value of the integrand, and the error is given by
\[
\epsilon_t = \ddot{Q} \Delta t^2 .
\]
With $\ddot{Q}$ we refer to a quantity related to the second time
derivative of $Q- \mathbf{u} \cdot \nabla A$. Its precise definition,
and that of $\bar{Q}$, will change depending on the particular
integration method and whether the simulation is Eulerian or
Lagrangian \cite{Idelsohn_2015}.  (In the latter case, the convective
term in $\ddot{Q}$ appears due to the errors in integrating the
particles' position, Eq. (\ref{eq:position_integration}) ).  These
details do not concern us, since we focus on the dependence on $\Delta
t$.

During a simulation with time span $T$, there will be
$N_t=T/\Delta t$ steps, hence the total accumulated error due to
time integration will be
\[
E_t = N_t ( \epsilon_t \Delta t) = \ddot{Q} T \Delta t^2 .
\]

For the error due to the linear spatial approximation of the FEM, a
similar result is obtained, with an error
\[
\epsilon_x = Q''  \Delta x^2  ,
\]
where $\Delta x$ is the average distance between the nodes used to
evaluate derivatives.  Again, $Q''$ is a quantity related to the
second space derivatives of $Q- \mathbf{u} \cdot \nabla A$. The
total accumulated error will therefore be
\[
E_x = N_t ( \epsilon_x \Delta t) = Q'' \Delta x^2 T.
\]


When increasing the resolution of a typical simulation, the value of
$\Delta x$ is usually tied to $\Delta t$ by a prescribed Courant
number
\[
\mathrm{Co} := \frac{u_0 \Delta t}{\Delta x},
\]
where $u_0$ is the maximum modulus of the velocity. The total error
can then be written as
\[
E = 
T
\left(
\ddot{Q}
  \Delta t^2  +
Q''
  \Delta x^2  
\right) = 
T  \Delta t^2 
\left(
\ddot{Q}
+
\frac{Q'' u_0^2}{\mathrm{Co}^2}
\right) .
\]
A convergence as $\Delta t^2$ is therefore expected for simple
methods, including Eulerian methods and Lagrangian methods such as
pFEM \cite{PFEM}, and linearly consistent SPH \cite{Liu_2003}.

%

If, on  the other hand, the  method involves a combination  of a fixed
mesh and  particles, another source of  error appears: the one  due to
projecting the fields from the mesh  onto the particles and back.  For
the projection from the mesh onto the particles we expect
\begin{equation}
\label{eq:eps_proj_mp}
\epsilon_\mathrm{mp}  = A'' H^2 ,
\end{equation}
where the  general $\Delta x$ is  given by $H$, the  mean mesh spacing
(see Fig \ref{fig:idea}).  At variance  with the previous errors, this
one is  independent of the  time spacing. The total  accumulated error
will be
\begin{equation}
\label{eq:E_proj_mp}
E_\mathrm{mp}  = A'' H^2 N_t = A'' H^2 \frac{T}{\Delta t}  =
A'' T \frac{ u_0^2 }{ \mathrm{Co}_H^2 } \Delta t ,
\end{equation}
where in the last equality the mesh Courant number is introduced, as
$\mathrm{Co}_H := u_0 \Delta t/H$. This error is seen to decrease
only as $\Delta t$, compared with $\Delta t^2$ for straight pFEM.

Similarly, for the projection from the particles onto the mesh one
expects
\begin{align}
  \label{eq:eps_proj_pm}
  \epsilon_\mathrm{mp}  &= A'' h^2 , \\
  \label{eq:E_proj_pm}
  E_\mathrm{pm} & = A'' T \frac{ u_0^2 }{ \mathrm{Co}_h^2 } \Delta t ,
\end{align}
where the particle Courant number is
$\mathrm{Co}_h := u_0 \Delta t/h$, with the mean interparticle
distance $h$ (see Fig \ref{fig:idea}).  In Ref. \cite{Idelsohn_2015}
this source of error is minimized by introducing many particles, thus
reducing $h$ and increasing $\mathrm{Co}_h$.

To summarize, the total error can be expected to be
\[
E =
T
\left(
\ddot{Q} \Delta t^2  +
 Q'' \Delta x^2  +
 A'' \frac{H^2}{\Delta t}  +
 A'' \frac{h^2}{\Delta t}  
\right) ,
\]
where an ambiguous $\Delta x$ is kept, to be fixed as $H$ or $h$
depending on the simulation. In terms of the Courant numbers, one obtains
\[
E=
T
\Delta t
\left(
  \Delta t
  \left[
    \ddot{Q} +    \frac{Q'' u_0^2}{ \mathrm{Co}^2}
  \right]
  +
  A'' u_0^2
  \left[
    \frac{1}{  \mathrm{Co}_H^2} +
    \frac{1}{  \mathrm{Co}_h^2} 
  \right]
\right) .
\]

Moreover, the ratio $m:=h/H$ is usually fixed as the simulation is
refined, resulting in a total error
\[
E=
T
\Delta t
\left(
  \Delta t
  \left[
    \ddot{Q} +    \frac{Q'' u_0^2}{ \mathrm{Co}^2}
  \right]
  +
  A''     \frac{ u_0^2 }{  \mathrm{Co}_h^2}
  \left[
    1+
    \frac{1}{ m^2}
  \right]
\right) ,
\]
which only decreases as $\Delta t$.

If, on the other hand, a quadratically consistent projection scheme
from particles to mesh is used, we will have instead of
Eqs. (\ref{eq:eps_proj_mp}) and Eqs. (\ref{eq:eps_proj_pm}),
\[
\epsilon_\mathrm{mp} = A''' H^3 \qquad \epsilon_\mathrm{pm} = A''' h^3 .
\]

We may then expect a total error
\[
E =
T
\left(
  \ddot{Q} \Delta t^2  +
  Q''      \Delta x^2  +
  A''' \frac{H^2}{\Delta t}  +
  A''' \frac{h^2}{\Delta t}  
\right) .
\]
In terms of the Courant numbers,
\[
E=
T
\Delta t^2
\left(
  \ddot{Q} +    \frac{Q'' u_0^2}{ \mathrm{Co}^2}
  +
  A''' u_0^3
  \left[
    \frac{1}{  \mathrm{Co}_H^3} +
    \frac{1}{  \mathrm{Co}_h^3} 
  \right]
\right) ,
\]
indeed restoring $\Delta t^2$ dependence.  A procedure to build such a
scheme is described on the next subsection.

\subsection{Shape functions}
\label{sec:shape}

Let us consider two sets of $N$ points in space as in
Fig. \ref{fig:idea}.  The points of the first set will be considered
``particles'' and have positions ${\bf r}_\mu$, with Greek index
labels.  The points of the second set will be considered ``mesh
nodes'' and have positions ${\bf r}_i$, with Latin index
labels. Associated to each set, we introduce two sets of shape
functions $\{\psi^\mathrm{p}_\mu(\bfr)\}$,
$\{\psi^\mathrm{m}_i(\bfr)\}$ that are used for interpolation.  Given
a field with value $\alpha_\mu$ at particle $\mu$, we construct the
interpolated field at an arbitrary point of space ${\bf r}$
\begin{equation}
  \label{eq:shape_particle}
  \alpha(\bfr) \doteq \sum_\mu \alpha_\mu \psi^\mathrm{p}_\mu(\bfr) .
\end{equation}
In a similar way, given a field value $\alpha_i$ at the mesh node $i$,
the interpolated field is given by
\begin{equation}
  \label{eq:shape_mesh}
  \alpha(\bfr) \doteq \sum_i \alpha_i \psi^\mathrm{m}_i(\bfr) .
\end{equation}
Mapping field values $\alpha_\mu$ at particles to field values
$\alpha_i$ at the mesh is given by
\begin{align}
  \label{p2m}
  \alpha_i \doteq \sum_\mu \alpha_\mu \psi^\mathrm{p}_\mu(\bfr_i) ,
\end{align}
while mapping from the  mesh to the particles is given by
\begin{equation}
  \label{m2p}
  \alpha_\mu \doteq \sum_i \alpha_i \psi^\mathrm{m}_i(\bfr_\mu) .  
\end{equation}
We will refer to the operation in (\ref{p2m}) as projection from
particles to mesh, and that in (\ref{m2p}) as projection from mesh to
particles, respectively. As explained in the Introduction, these
projection procedures are simply interpolation.

A particular simple choice for $\psi^\mathrm{p}_\mu({\bf r})$ and
$\psi^\mathrm{m}_i(\bfr) $ is the linear finite element shape
functions (FE).  Since the particles form a non-structured mesh, the
Delaunay triangulation is computed in order to build these FE
functions.  As the preceding discussion predicts, and our results
below confirm, this choice results in projection methods that perform
quite poorly.  In this work, therefore, quadratically consistent
interpolating shape functions (QCSF) are considered \cite{QCSF}.  The
procedure to compute these functions is explained briefly in Appendix
\ref{sec:quad}.  The idea is to use the original linear FE shape
functions \emph{and their products} in order to build an extended set
of shape functions that complies with quadratic consistency.

The problem of interpolation is related but different from the problem
of discretization of the continuum equations.  Here we will use a
standard Galerkin method for the discretization, with the same set of
shape functions used for interpolation. A number of different methods
are considered and compared in the present work.  They are summarized
in Table \ref{table:methods} depending on the set of Galerkin basis
and interpolating functions used.  For example, the ``particle Finite
Element Method'' (partFEM) is a Lagrangian method introduced in
\cite{PFEM}, where there is no projection to a mesh and linear finite
element shape functions are used.  Our partFEM is simply a particular
implementation of the pFEM method of \cite{PFEM}, differing on details
like using CGAL libraries, for example.  If projection from particles
to mesh and back is considered, the name ``projection Finite Element
Method'' (projFEM) will be used.  If QCSF functions are used for both,
the interpolation from the mesh to the particles, and for the inverse
procedure, the method projFEMq described below results.  However, we
have also considered the case in which QCSF are used for projection
from the mesh to the particles, but only linear FEs for the inverse
procedure (method projFEM6 below).  This is considered since it is
computationally simpler and faster, and in order to connect with
previous work in Ref. \cite{Idelsohn_2015}.  A more precise discussion
of the methods summarized in the Table will be given in Section
\ref{sec:zalesak}.  Many other combinations have been explored, but as
will be explained, the additional computational cost often results in
a negligible improvement, in terms of accuracy vs CPU time.

\begin{table} \centering
  \begin{tabular}{|c|c|c|c|c|c|}
    \hline
     method's        &  \multicolumn{2}{c|}{Galerkin functions} &  \multicolumn{2}{c|}{Interpolating functions} &  \\
     name              &    particle         &  mesh             &     particle           &  mesh              & Similar to \\
    \hline
    \hline
    partFEM           &  FE                 & n/a               & n/a                    & n/a                & pFEM \cite{PFEM} \\
    projFEM6       &  FE                 & QCSF             & FE                     & QCSF              &   pFEM-2 \cite{Idelsohn_2015} \\
    projFEMq       &  QCSF              & QCSF             & QCSF                  & QCSF              &  \\
    \hline
  \end{tabular}
  \caption{Features of the methods considered in this work, left column, and their features. FE: linear finite element shape functions.
    QCSF: quadratically consistent shape functions. Last column: previous works with similar methods.
    \label{table:methods}}
\end{table}

\section{Results}
\label{sec:results}

\subsection{Rigid rotation of Zalesak's disk}
\label{sec:zalesak}

Let us consider a region with a color field $\alpha$ that has a value
of $1$ for points inside a domain and $0$ for points outside, and
which is simply advected:
\begin{equation}
  \frac{d \alpha}{d t} = 0 ,
\end{equation}
where the time derivative is the convective derivative:
$d \alpha /dt = \partial \alpha / \partial t + (\mathbf{u} \cdot
\nabla) \alpha$.

The domain is a circle with a slot. The circle's radius is given a
value of $0.5$, while the slot was a width of $1/6$, and a height of
$5/6$. The simulation box is a $(-1,1)\times(-1,1)$ square, and the
number of nodes is set to $60 \times 60$, so that the mesh spacing is
$H=2/60=1/30$, the same value as in \cite{Idelsohn_2015}.  The time
step is $\Delta t=0.01$, which corresponds to
$\mathrm{Co}_H := u \Delta t /H \approx 0.94$, for nodes on the rim of
the disk.

We assume that the velocity field is given by a pure rotation for
points within a radius of $0.9$:
\begin{align}
  u_x &= -\omega y\\
  u_y &=  \omega x ,
\end{align}
where $\omega=2\pi/\tau$, and the period of rotation is set to
$\tau=1$.  Periodic boundary conditions are used in this simulation,
but this fact is not really important since the only region that is
actually moved is the region within the circle of radius $0.9$.

Since the field is just advected, it makes little sense to project
from and onto the mesh at every time-step. With no projection, the
shape just rotates, with some distortion due to the time integration,
and of course there is no diffusion (every particle has a value of
either $0$ or $1$).  Nevertheless, in order to benchmark our method
the $\alpha$ field is projected from the mesh to the particles and
vice versa at every time step. The precise procedure is:
\begin{enumerate}
\item Initialize field $\alpha$ on the mesh.
\item \label{enum:zal_init} For the mesh only: determine the
  coefficients needed for the evaluation of the shape functions
  $\psi^\mathrm{m}_i(\bfr)$. Since quadratic shape functions are used
  for the mesh, this requires the geometrical coefficients $A_{ij}$
  \cite{QCSF}.
\item \label{enum:zal_first2} Project the value of the fields from the
  mesh onto the particles, using Eq. (\ref{m2p}).
\item Move the particles one step according to the midpoint velocity
  field.
\item For the particles only: determine the coefficients needed for
  the evaluation of the shape functions $\psi^\mathrm{p}_\mu(\bfr) $.
  If quadratic shape functions are used for the particles, this
  requires the geometrical coefficients $A_{\mu\nu}$.
\item Project the values of the fields from the particles back onto
  the mesh, using Eq. (\ref{p2m}).
\item Go to \ref{enum:zal_first2} until the desired number of steps.
\end{enumerate}

\begin{figure}
  \centering
  \begin{minipage}{0.3\textwidth}
      \includegraphics[width=\textwidth]{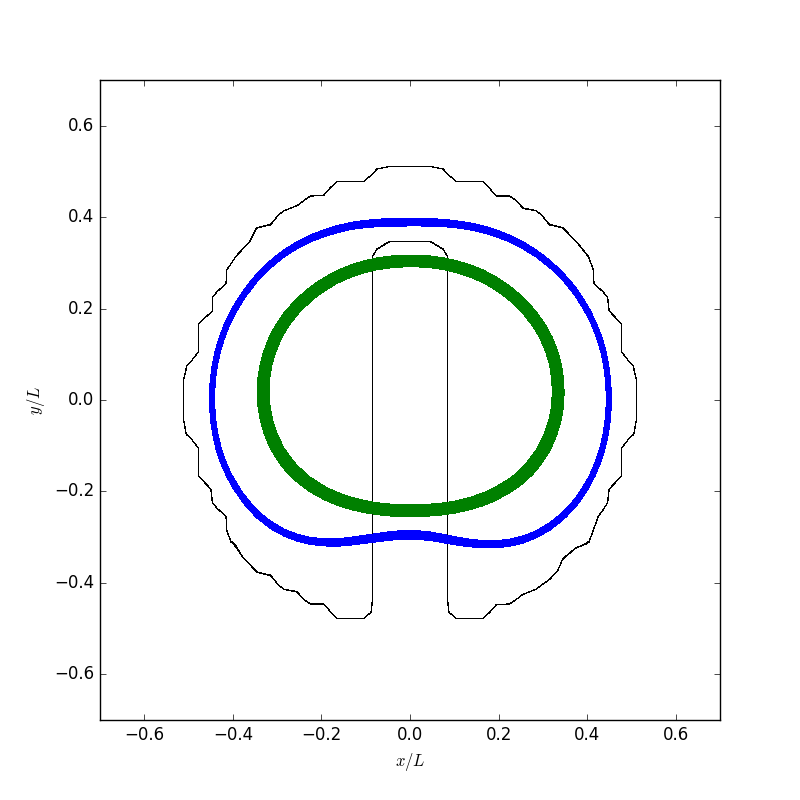}
  \end{minipage}
  \quad
  \begin{minipage}{0.3\textwidth}
      \includegraphics[width=\textwidth]{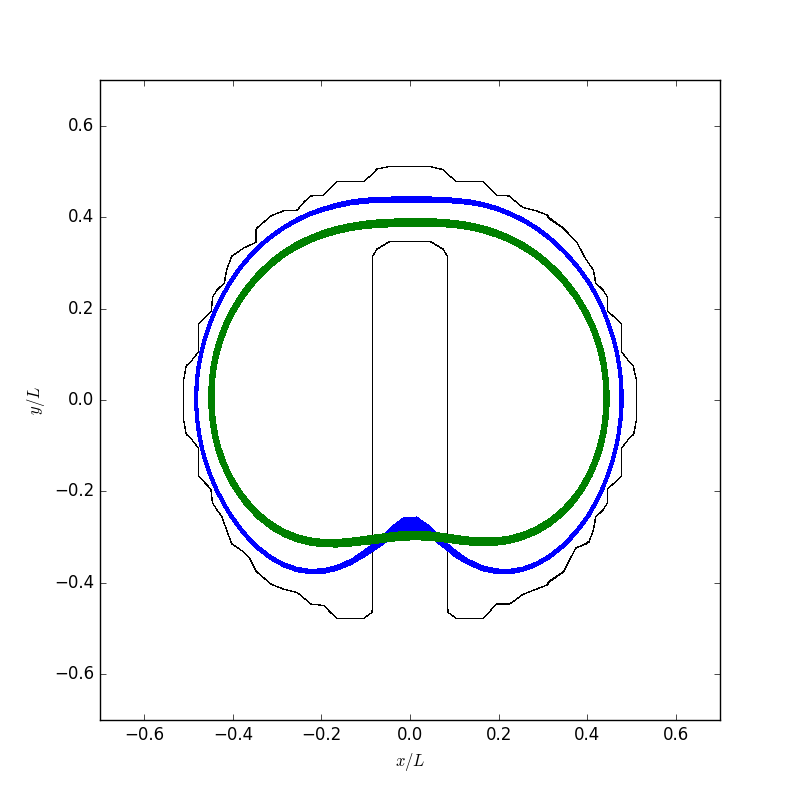}
  \end{minipage}
  \quad
  \begin{minipage}{0.3\textwidth}
      \includegraphics[width=\textwidth]{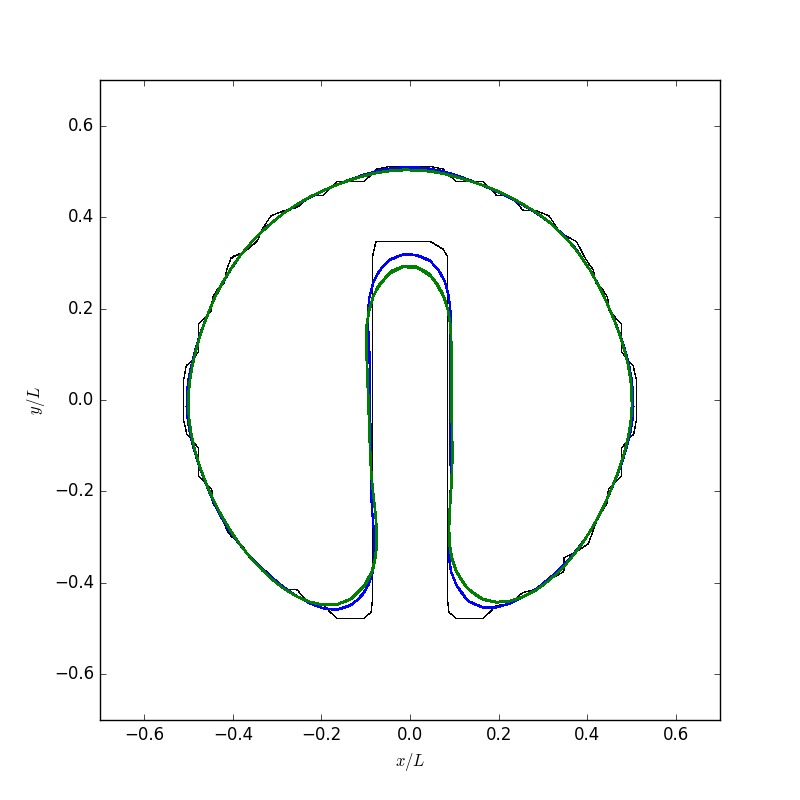}
  \end{minipage} \\
  \begin{minipage}{0.3\textwidth}
      \includegraphics[width=\textwidth]{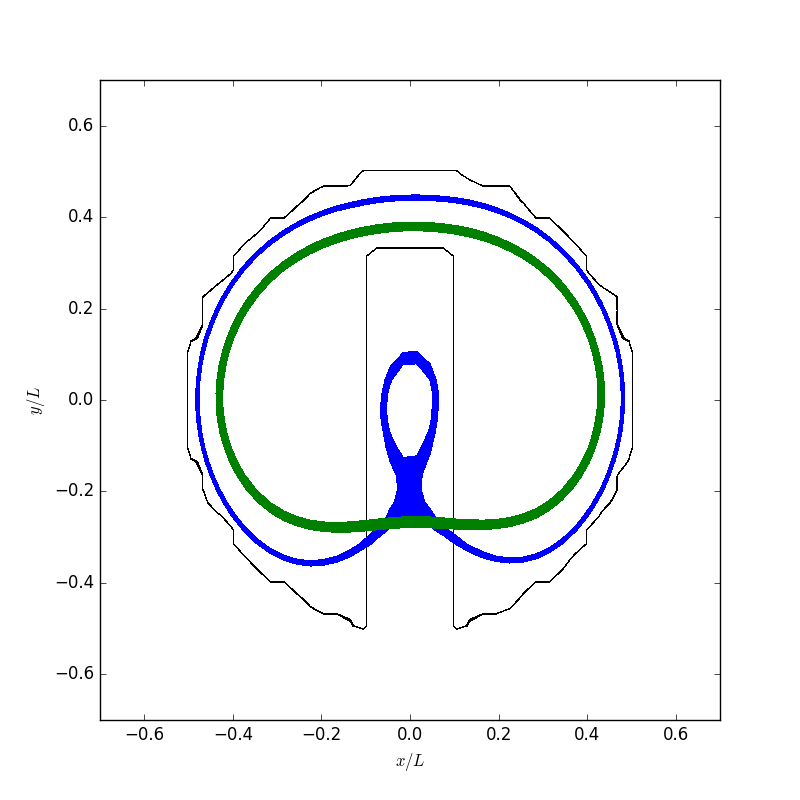}
  \end{minipage}
  \quad
  \begin{minipage}{0.3\textwidth}
      \includegraphics[width=\textwidth]{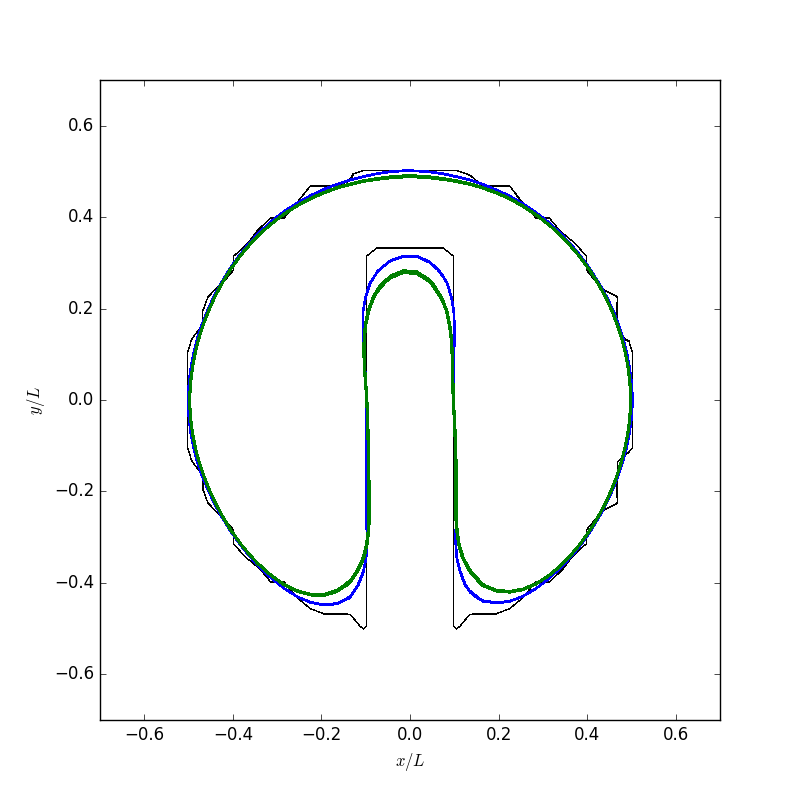}
  \end{minipage}
  \quad
  \begin{minipage}{0.3\textwidth}
      \includegraphics[width=\textwidth]{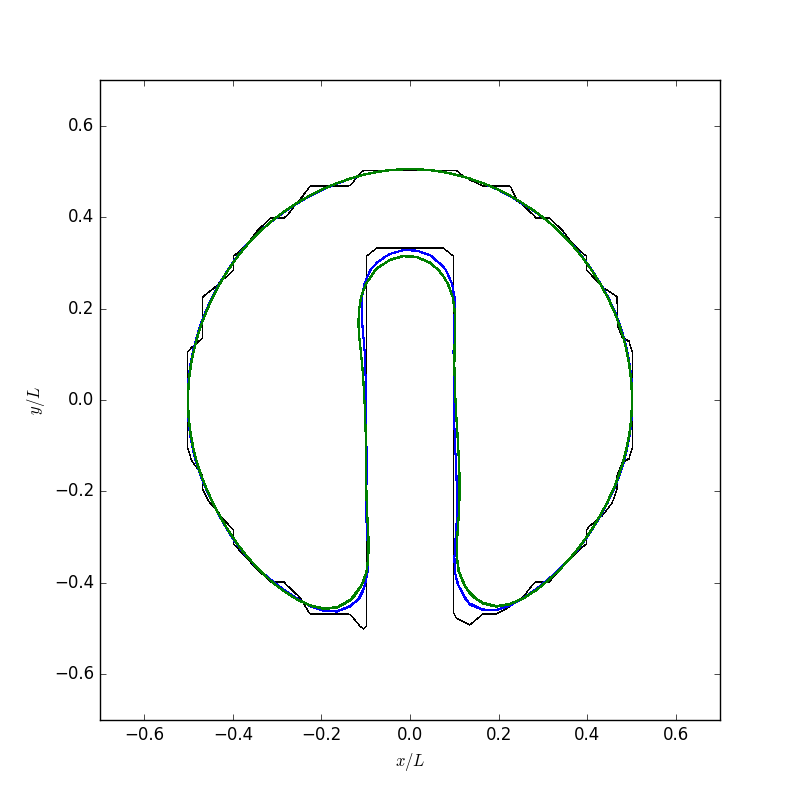}
  \end{minipage}
  \caption{Results for the rotation of Zalesak's disk.  Isocontours
    for $\alpha \in ( 0.49 , 0.51 )$. Initial field in black, after
    one rotation in blue, after two in green. Top row: ``coarse''
    simulations, same number of particles as mesh nodes ( $h=H$ ).
    Bottom row: ``fine'' simulations, six particles per mesh nodes.
    Left column: linear FEs for mesh and particles. Middle: QCSF for
    the mesh, linear for the particles.  Right: QCSF for both mesh and
    particles.\label{fig:zalesak}}
\end{figure}

Results are given in Fig. \ref{fig:zalesak}, showing contour plots for
values between $0.49$ and $0.51$ of the $\alpha$ field on the mesh
nodes.  We include the initial contour, the contour after one
revolution, $T=\tau$, and after two revolutions, $T=2\tau$. On the top
row, results are shown for a ``coarse'' simulation, with the same
number of particles as nodes, $h=H$.  First, it can be seen that a
simple, linear method (linear FEs for both the mesh and the particles)
does not work at all for the coarse simulation (top left): the contour
spreads and shrinks, even disappearing on the second revolution.  If
QCSF are used for the mesh, the simulation improves, but the slot is
quickly lost (top middle).  However, by including QCSF for both, mesh
and particles, the simulation turns out to be satisfactory (top
right).

The lower row shows results for a ``fine'' simulation, with $6$
particles per mesh node, or $h=H/\sqrt{6}$.  Results for a linear
method (bottom left) are quite poor, but with QCSF the simulation
turns out to be quite satisfactory (bottom middle). This fact confirms
the findings of Ref. \cite{Idelsohn_2015}, in which a large number of
particles per node is used.  Finally, if the particle shape functions
also satisfy quadratic consistency the contours (bottom right) are
much improved. Note that the computational costs are quite high for
the latter, as we discuss later on.

These results confirm our previous analysis on how the consistency of
the spatial approximation should have a direct impact on the overall
error of the simulation. From now on, the focus will therefore be on
two particular implementations of projFEM: \emph{projFEMq},
corresponding to the top right of Fig.  \ref{fig:zalesak}, with QCSF
both for the mesh and the particles, and as many particles as mesh
nodes --- and \emph{projFEM6}, corresponding to bottom middle of
Fig. \ref{fig:zalesak}, with quadratic mesh shape functions, linear
particle shape functions, and $6$ times more particles than mesh
nodes.

\begin{figure}
  \centering
  \begin{minipage}{0.45\textwidth}
      \includegraphics[width=\textwidth]{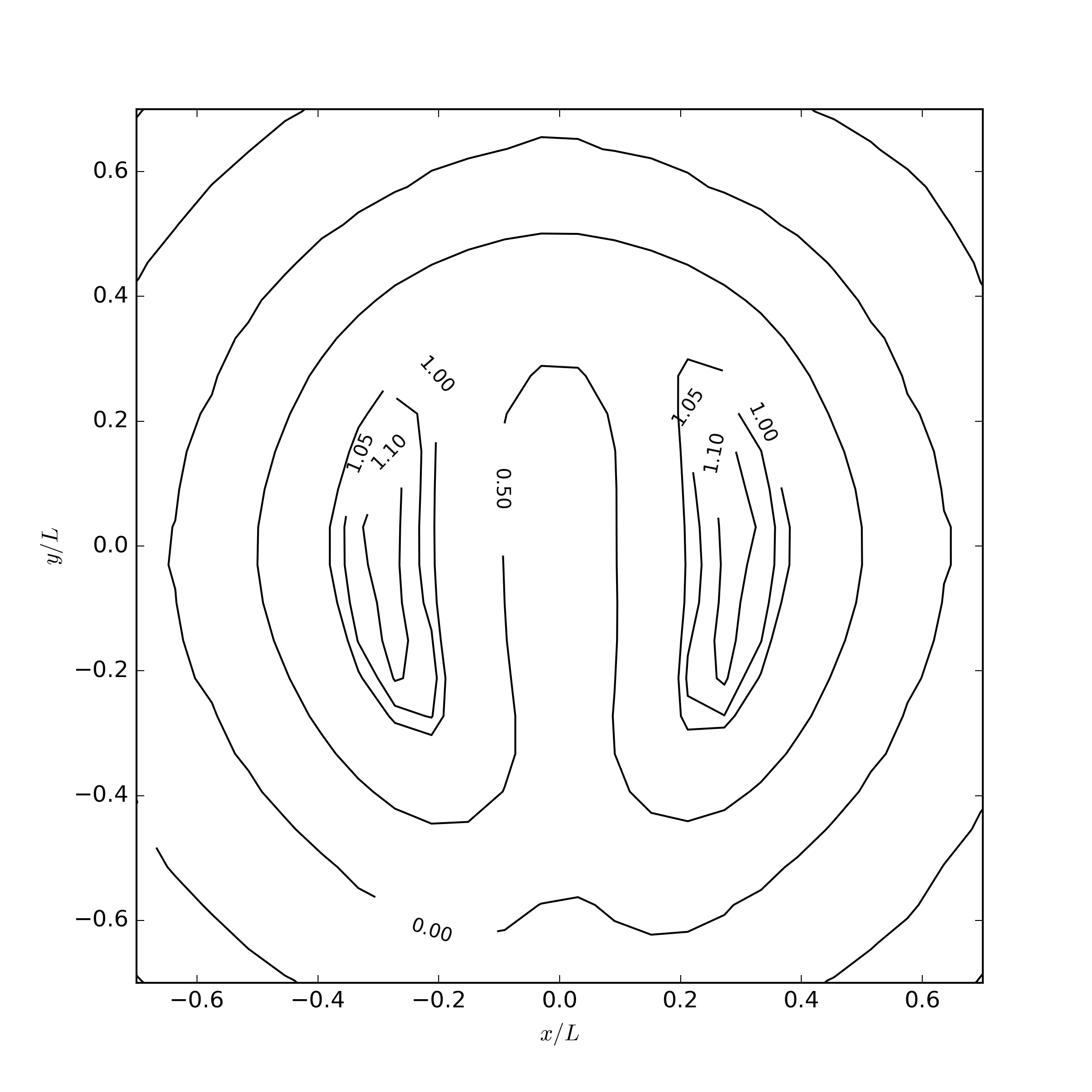}
  \end{minipage}
  \quad
  \begin{minipage}{0.45\textwidth}
      \includegraphics[width=\textwidth]{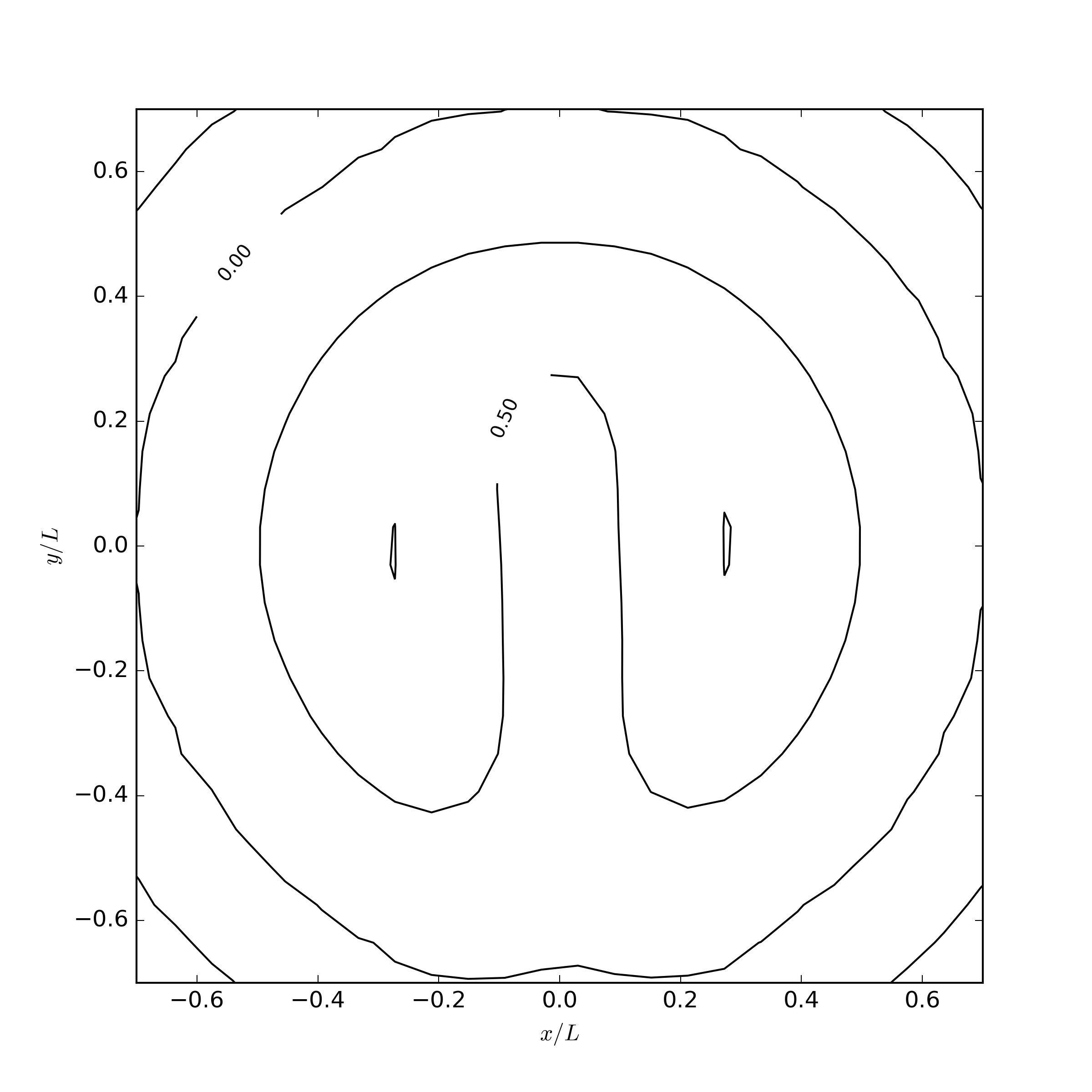}
  \end{minipage}
  \caption{Results for the rotation of Zalesak's disk.  Isocontours
    for $\alpha \in ( 0, 0.5 , 1 , 1.05 , 1.1 )$ after two
    rotations. Left: projFEMq. Right: projFEM6. \label{fig:zalesak2}}
\end{figure}

On the other hand, usage of the extended functional set results in
Gibbs phenomena for this discontinuous field. Fig. \ref{fig:zalesak2}
shows detailed contours for projFEMq and projFEM6 after two rotations.
The isocontour for $\alpha=0.5$ is shown again, but the $\alpha=0$
contour reveals a corona of negative values that is more spread out in
projFEM6. Values of $\alpha$ as high as $1.1$ are seen in the inner
regions for projFEMq. These high values are not obtained for projFEM6,
but in this case the high values are greatly eroded, with just two
small slits for $\alpha=1$.  This limitation is not surprising, since
the extended functions have negative regions, as is apparent in
condition (\ref{eq:A0}) of Appendix \ref{sec:quad}.

\subsection{Taylor-Green vortex sheet}
\label{sec:TG}

This is an analytic solution to the Navier-Stokes equations for an
incompressible Newtonian fluid:
\begin{equation}
  \frac{d \mathbf{u}}{d t} =  - \nabla p +  \nu \nabla^2 \mathbf{u}.
\end{equation}

The solution in periodic boundary conditions is described by the
velocity field
\begin{align}
\label{eq:TG_vel}
  \mathbf{u}_x &=  f(t) \sin (k x) \cos (k y) \\
  \mathbf{u}_y &= -f(t) \cos (k x) \sin (k y) ,\\
\end{align}
with $k=2\pi/L$, and the pressure field
\begin{equation}
  \frac{p}{\rho} = \frac1 4 f(t)^2
  \left[
    \cos (2 k x)  + \cos (2 k y)
  \right] ,
\end{equation}
where
\[
f(t)=u_0 \exp\left( - 8 \pi^2 t^* /\mathrm{Re} \right) .
\]
The Reynolds number is defined as Re$:=u_0 L / \nu$, and dimensionless
time is $t^* := t u_0/L$. We set $ u_0=1$, $L=2$, and $\nu=0.01$,
setting a Reynolds number Re=$200$.

For the numerical solution of the Navier-Stokes equation, a standard
splitting approach is used \cite{codina_2001}. For the partFEM case,
the algorithm employed is:
\begin{enumerate}
\item Initialize the velocity field $ \bfu^\mathrm{m}_{t=0}$ on the
  mesh, from Eq. (\ref{eq:TG_vel}).
\item \label{enum:first} Define a midpoint velocity $\bfu_{n+1/2}
  \approx \bfu_{n}$.
\item \label{enum:first2} Move the particles a half step according
  to the midpoint velocity field:
  $
  \bfr_{n+1/2} =   \bfr_{n} + \frac{\Delta t}{2} \bfu_{n+1/2} .
  $
\item Build a Delaunay triangulation from the position of the
  particles.  Determine the coefficients needed for the evaluation of
  linear FEM shape functions and their integrals.  Determine the
  matrices needed for the FEM calculation.
\item Calculate  an intermediate  velocity field  from the  equation $
  \bfu^* = \bfu_{n} + \frac{\Delta t}{2} \mu \nabla^2 \bfu^* .  $
\item Solve the Poisson pressure equation
  $
  \nabla^2  p_{n+1/2} = \frac{2}{\Delta t} \nabla\cdot\bfu^*  .
  $
\item Calculate new midpoint velocities
  $\bfu_{n+1/2} = \bfu^* - \frac{\Delta t}{2}  \nabla p_{n+1/2}  $.
\item Go to \ref{enum:first2} until convergence in positions (i.e. the
  new positions are within a distance threshold from the previous
  iteration).
\item Move the particles a whole step:
  $
  \bfr_{n+1} =   \bfr_{n} + \bfu_{n+1/2} \Delta t
  $.
\item Go to  \ref{enum:first} until the end
\end{enumerate}
Each of these operations is of course carried out on each of the
particles, therefore we should have written $\bfu_{i,n+1/2} =
\bfu_{i,n}$, etc, but the $i$ index is dropped for the sake of
simplicity.



For projFEM, the procedure is very similar, with the space derivatives
being calculated on the mesh (namely: solving for $\bfu^*$, solving
for the pressure, and evaluating its gradient).  The procedure is as
follows, where a ``$\mathrm{m}$'' superscript is used to distinguish
the mesh velocity field, and ``$\mathrm{p}$'' for the one on the
particles.
\begin{enumerate}
\item \label{enum:proj_init} Build the Delaunay triangulation for the
  mesh. Determine the coefficients needed for the evaluation of QCSF
  shape functions and their integrals.  Determine the matrices needed
  for the Galerkin calculation.
\item Initialize the velocity field $ \bfu^\mathrm{m}_{t=0}$ on the
  mesh, from Eq. (\ref{eq:TG_vel}).
\item \label{enum:proj_first} At time-step $n$, define a midpoint mesh
  velocity
  $
  \bfu^\mathrm{m}_{n+1/2} \approx \bfu^\mathrm{m}_{n}
  $.
\item \label{enum:proj_first2} Project the midpoint velocity onto
  the particles $\bfu^\mathrm{m}_{n+1/2} \rightarrow
  \bfu^\mathrm{p}_{n+1/2}$.
\item Move the particles a half step according to the midpoint
  velocity field:
  $ \bfr_{n+1/2} = \bfr_{n} +
  \frac{\Delta t}{2} \bfu^\mathrm{p}_{n+1/2} .
  $
\item Build the Delaunay triangulation for the particles. Determine
  the coefficients needed for the evaluation of either linear FEM
  shape functions (projFEM6) or QCSF (projFEMq).
\item Project the previous velocity onto the mesh,
  $\bfu^\mathrm{p}_{n} \rightarrow \bfu^\mathrm{m}_{n}$.
\item Calculate an intermediate mesh velocity field from the equation
  $ \bfu^* = \bfu^\mathrm{m}_{n} + \frac{\Delta t}{2} \mu \nabla^2 \bfu^* .  $
\item Solve the Poisson pressure equation on the mesh
  $
  \nabla^2  p_{n+1/2} = \frac{2}{\Delta t} \nabla\cdot\bfu^*  .
  $
\item Calculate new midpoint mesh velocities
  $\bfu^\mathrm{m}_{n+1/2} = \bfu^* - \frac{\Delta t}{2}  \nabla p_{n+1/2}  $.
\item  Go to  \ref{enum:proj_first2}  until  convergence in  positions
  (i.e.  the new  positions of  the  particles are  within a  distance
  threshold from the previous iteration).
\item Move the particles a whole step:
  $
  \bfr_{n+1} =   \bfr_{n} + \bfu^\mathrm{p}_{n+1/2} \Delta t
  $. Calculate new velocities:
  $
  \bfu^\mathrm{p}_{n+1} =  2 \bfu^\mathrm{p}_{n+1/2} - \bfu^\mathrm{p}_{n}
  $
\item Build the Delaunay triangulation for the particles. Determine
  the coefficients needed for the evaluation of either linear FEM
  shape functions (projFEM6) or QCSF (projFEMq).
\item Project the new velocities onto the mesh,
  $\bfu^\mathrm{p}_{n+1} \rightarrow \bfu^\mathrm{m}_{n+1}$.
\item Go to  \ref{enum:first} until the desired number of steps.
\end{enumerate}

An obvious computational advantage compared to the previous,
all-particle, partFEM is that the mesh shape functions and the
matrices needed for the spatial differential equations are calculated,
and treated, only once in the course of the simulation, at step
\ref{enum:proj_init}.  However, the particle shape functions are
calculated at every step since they are still needed for the
projection procedure.


The projFEM6 method is very similar to the pFEM2 methods of Ref.
\cite{Idelsohn_2015}, with fully explicit integration of the positions
and fully implicit integration on the mesh (a value of $\theta=1$ in
pFEM2's notation).  The main difference is that our integrators are
simple half-step integrators, while in pFEM2 the values of the fields
are integrated following the streamlines.  The projFEMq method further
deviates from pFEM2 by including QCSF (see Table \ref{table:methods}).

\begin{figure}
  \centering
  \begin{minipage}{0.42\textwidth}
    \includegraphics[width=\textwidth]{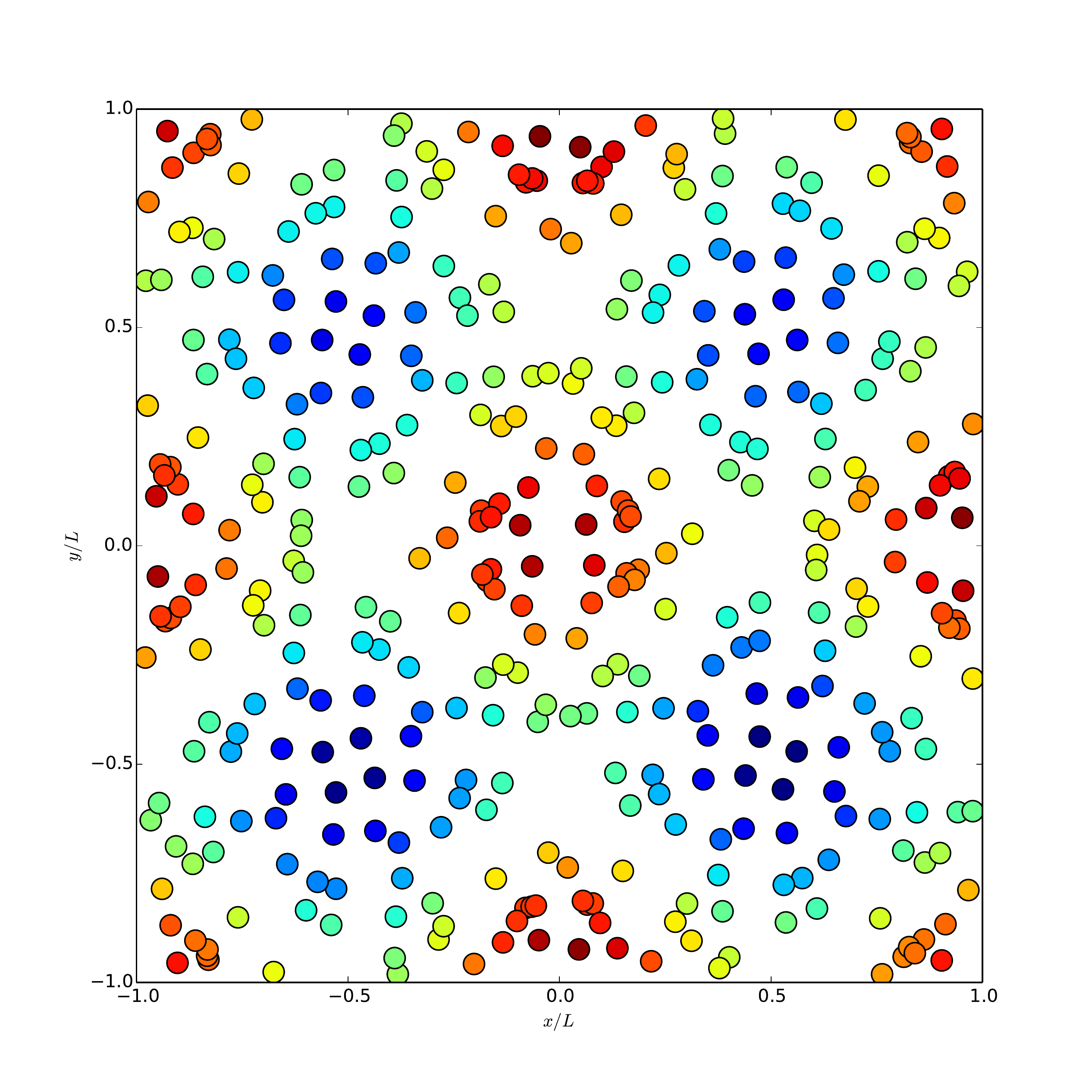}
  \end{minipage}
  \quad
  \begin{minipage}{0.42\textwidth}
    \includegraphics[width=\textwidth]{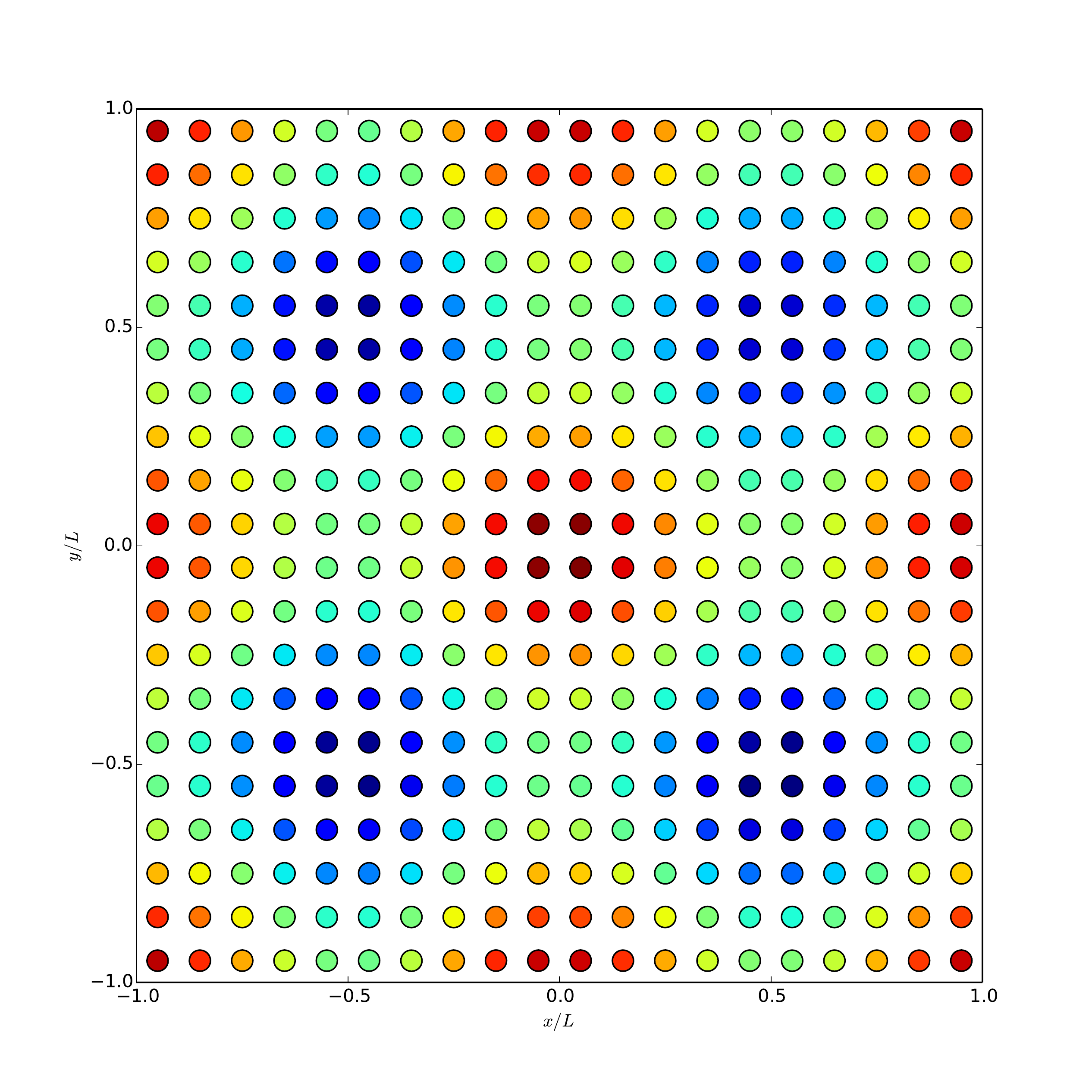}
  \end{minipage}
  \\
  \begin{minipage}{0.42\textwidth}
    \includegraphics[width=\textwidth]{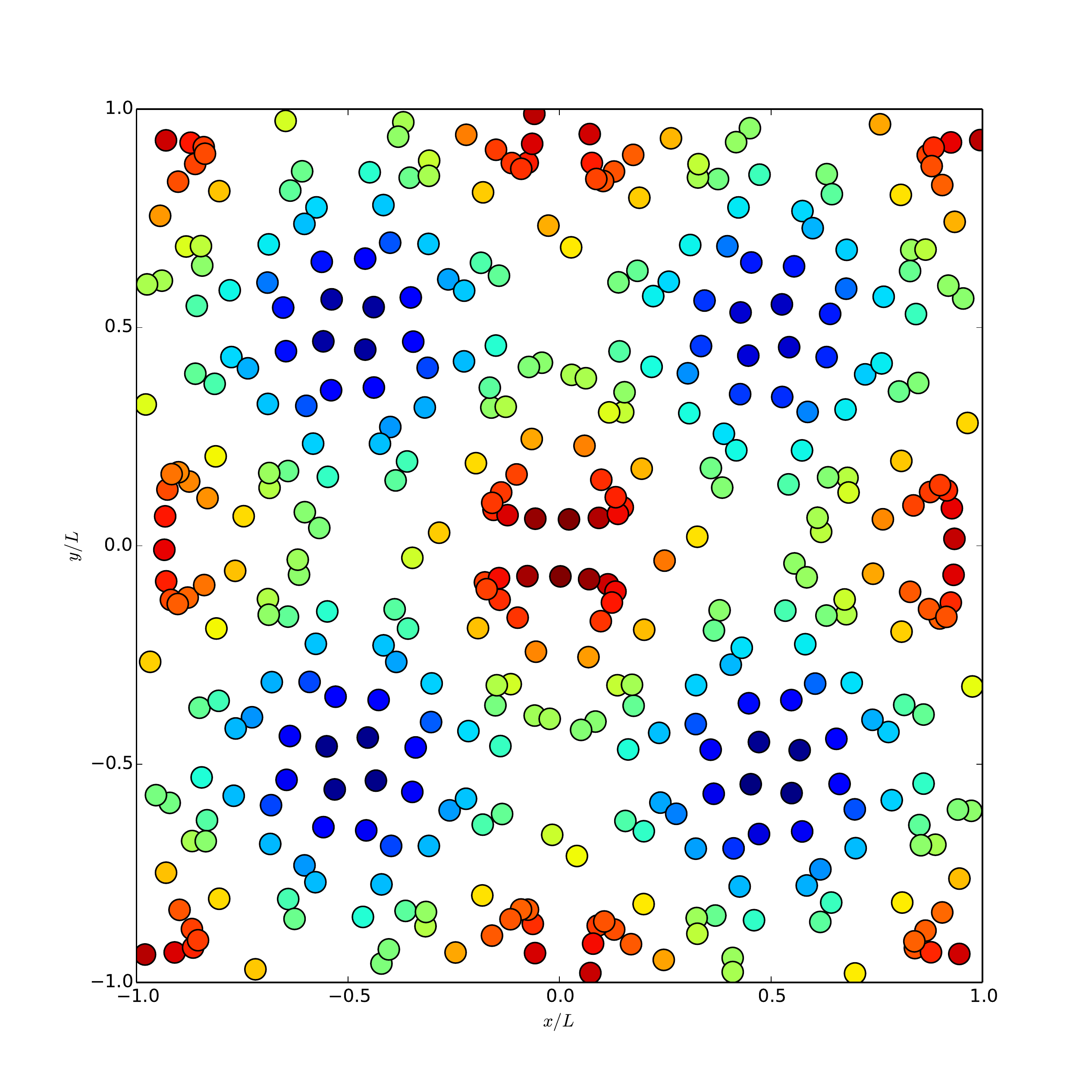}
  \end{minipage}
  \quad
  \begin{minipage}{0.42\textwidth}
    \includegraphics[width=\textwidth]{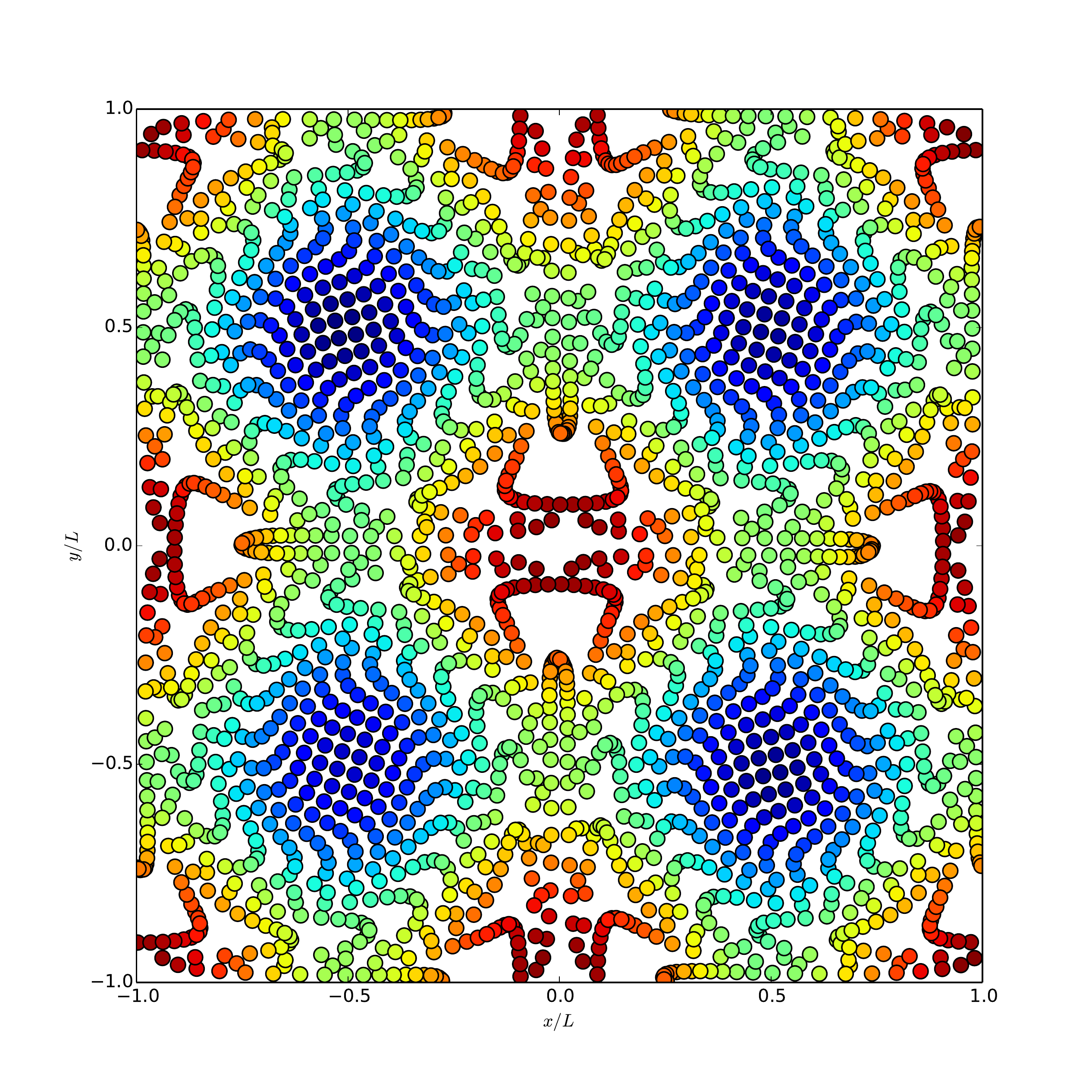}
  \end{minipage}
  \caption{Results for the pressure field at $T^*=1$. From top left:
    partFEM, projFEMq (field on the mesh), projFEMq (field on the
    particles), projFEM6 (field on the particles).\label{fig:TG}}
\end{figure}

In Fig. \ref{fig:TG} we show a snapshot of a simulation that has
started from a regular arrangement, for a number of particles
$N=20 \times 20$, a time-step of $\Delta t^*=0.025$ (corresponding to
a Courant number of Co$_H=0.5$), at a reduced time $T^*=1$. The top
left figure shows the particles' position and pressure field for the
partFEM simulation.  At its right, the pressure field for the mesh in
a projFEMq simulation is plotted.  The corresponding particles are
plotted at the bottom left, with the same color code (actually, the
pressure field is calculated only on the mesh, but here it has been
interpolated on the particles for visualization purposes).  Finally,
the figure on its right shows the particles for an projFEM6 simulation
(the mesh result is visually very similar to the one above and is
therefore not shown).

In order to quantify the accuracy of the different methods, 
the relative $L_2$ distance between a given field $\phi$ obtained by
simulation and its exact value is computed:
\begin{equation}
\label{eq:L2_error}
L_2:=
\sqrt{%
  \frac{
    \sum_{i=1}^N 
    V_i
    \left|
      \phi(\bfr_i) - \phi_i
    \right|^2
  }{
    \sum_{i=1}^N 
    V_i
    |\phi(\bfr_i)|^2
  }
} ,
\end{equation}
where $\phi(x_i)$ is the exact solution at particle $i$ and $\phi_i$
the computed one on that particle position. The volume of particle $i$
is defined as $V_i:=\int \psi_i(\bfr) d\bfr$. The expression results
from the discretization of the $L_2$ distance between two functions,
$L_2:=\int (f(\bfr)-g(\bfr))^2 d\bfr$.  The same measure may be
evaluated for mesh nodes, with very similar results. For vector fields
this distance is defined with $|\cdots|^2$ meaning the squared vector
modulus.

\begin{figure}
  \centering 
  \includegraphics[width=0.8\textwidth]{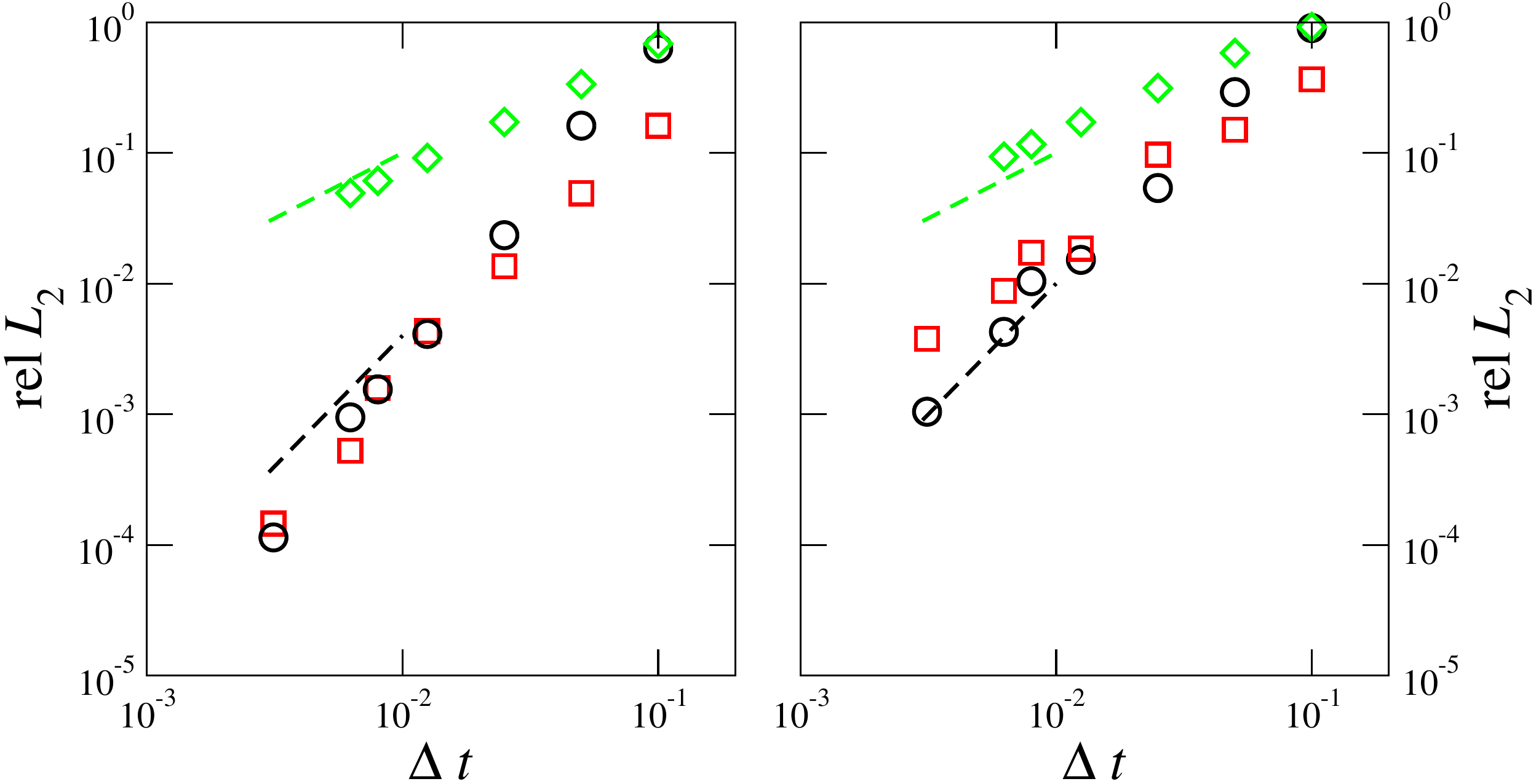}
  \caption{%
    $L_2$ error of the velocity (left) and pressure (right) fields at
    time $T^*=1$ versus time step $\Delta t$.  Black circles: projFEMq
    method, red squares: partFEM, green diamonds: projFEM6.  Dashed lines
    are power-laws plotted for convenience.
     Left: $\Delta t^2$ (lower black line), $\Delta t$ (upper green line).
    Right: $\Delta t^2$ (lower black line), $\Delta t$ (upper green line).
    \label{fig:L2_vs_Dt}
  }
\end{figure}

This error is expected to start at a very low value and increase
approximately linearly as the simulation proceeds. In order to compare
between methods, in Fig. \ref{fig:L2_vs_Dt} the value of this error at
$T^*=1$ is plotted, for the velocity and pressure fields.  At this
time, $f(T)=\exp(-8\pi^2 / 200 )=0.67 $, so that the velocity field
should have decreased to about $67\%$ of its initial value, and the
pressure, $45\%$.

The error for the velocity field (left subfigure) is seen to decrease
with $\Delta t$. The value of $h$ (and, for projFEM methods, $H$) also
decreases as $\Delta t$, in order to fix a Courant number of
Co$_H=0.5$.  As expected, the order of convergence of the error agree
with a $\Delta t^2$ power low both for partFEM and projFEMq. The error
of projFEM6 is seen to vanish only as $\Delta t^1$.  The error for the
pressure (right subfigure) vanishes roughly similarly: as $\Delta t^2$
for partFEM and projFEMq, and as $\Delta t^1$ for projFEM6.  (This
error is evaluated at the nodes, since the pressure field is obtained
on them and need not be projected onto the particles.)


A relevant question when comparing different methods is their efficiency
in terms of CPU time. By plotting the error as a function of CPU time
one may discriminate between a straightforward method that performs
quickly, and a more sophisticated one that needs more computational
resources. Results for the latter clearly depend on the machine, but
if we employed a faster one, it is likely that the CPU times of each
simulation run will be sped up by roughly the same factor. This would
simply result in a horizontal translation of all the curves in a
logarithmic scale. Results do depend on the particular linear algebra
algorithm used, details can be found in Appendix \ref{sec:numerical}.

In Fig. \ref{fig:L2_vs_time} the relative $L_2$ error at simulation
time $T^*=1$ is plotted as a function of the CPU time
$T_\mathrm{CPU}$.
Regarding the velocity field (left subfigure in
Fig. \ref{fig:L2_vs_Dt}), the convergence for partFEM is as
$T_\mathrm{CPU}^{-0.7}$, but is slightly superior, as
$T_\mathrm{CPU}^{-0.85}$, for projFEMq, due to the latter using direct
methods for numerical algebra. Nevertheless, partFEM may be preferable
at lower resolutions, as the crossover takes place at the finest,
longest, simulations.  The convergence of projFEM6 is a rather poor
$T_\mathrm{CPU}^{-0.3}$. For the pressure (right subfigure), the
convergence for partFEM is about $T_\mathrm{CPU}^{-0.4}$, for projFEMq
it is $T_\mathrm{CPU}^{-0.7}$, and for projFEM6 it is
$T_\mathrm{CPU}^{-0.3}$. For this quantity the convergence of projFEMq
is clearly superior, and the crossover with partFEM occurs at
intermediate resolutions, due to the quite different rates of
convergence.

Therefore, one may conclude from these observations that the higher
computational cost of a projFEM method can be compensated at high
resolutions by its higher accuracy. When running the simulations it is
apparent how the partFEM method proceeds at a steady pace, while the
projFEM methods employs some time in the initial set-up of the
matrices, and then starts, taking more time at each time step, but
producing more accurate results.

\begin{figure}
  \centering
  \includegraphics[width=0.8\textwidth]{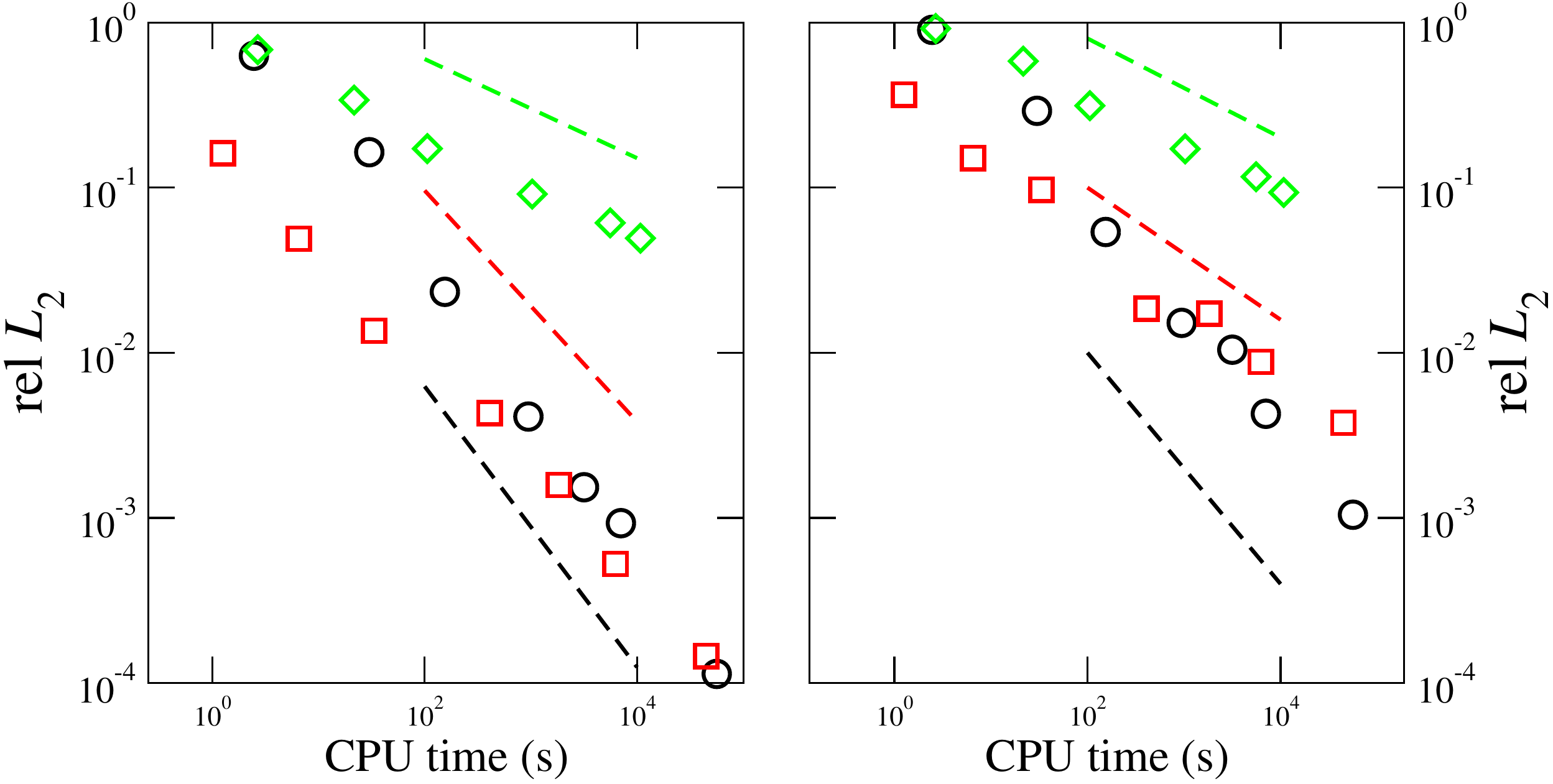}
  \caption{%
    $L_2$ error of the velocity (left) and pressure (right) fields at
    time $T^*=1$ versus CPU time.  Black circles: projFEMq method, red
    squares: partFEM, green diamonds: projFEM6.  Dashed lines are
    power-laws plotted for convenience.  Left: $T_\mathrm{CPU}^{-0.85}$
    (lower black line), $T_\mathrm{CPU}^{-0.7}$ (middle red line),
    $T_\mathrm{CPU}^{-0.3}$ (upper green line).  Right:
    $T_\mathrm{CPU}^{-0.7}$ (lower black line), $T_\mathrm{CPU}^{-0.4}$
    (middle red line), $T_\mathrm{CPU}^{-0.3}$ (upper green line).
    \label{fig:L2_vs_time}
  }
\end{figure}

\subsection{Rayleigh-Taylor instability}
\label{sec:RT}

In this section, we qualitatively compare the behavior of the methods
discussed, for the numerical solution of the Rayleigh-Taylor
instability, including results from OpenFOAM.  This well-known
benchmark case consists of two phases in a gravitational field, with
the denser layer on top.  The interface is perturbed initially, and
plumes develop.  As we do not want to consider wall boundary
conditions in this work, periodic boundary conditions are employed,
and a ``gravity'' of the form
\[
\vec{f}_\mathrm{g} = - g ( 2 \alpha - 1 ) \vec{u}_y ,
\]
acts on each particle, where $\alpha$ is a color function with values
$1$ for the heavier phase and $0$ for the lighter phase.  In this way,
heavier particles are pulled downwards and lighter particles,
upwards. The density is kept constant, and the value of $g$ is set to
$1$ for simplicity.

The CGAL libraries used in the Delaunay construction currently
implement only square periodic boundary conditions, therefore a square
simulation cell $(-1,1) \times (-1,1) $ is considered, instead of the
usual $1:4$ aspect ratio.  The initial interface is perturbed by a
cosine shape: $\eta = -0.05 \cos(4 \pi x )$, giving rise to four
identical plumes (or similar, for a randomly perturbed initial
distribution).  We set $\nu= 3.53\times 10^{-4} $, in order to set a
Reynolds number of Re$:=\sqrt{2L\sqrt{g L} }/\nu=1000$.

\begin{figure}
  \centering \includegraphics[width=0.8\textwidth]{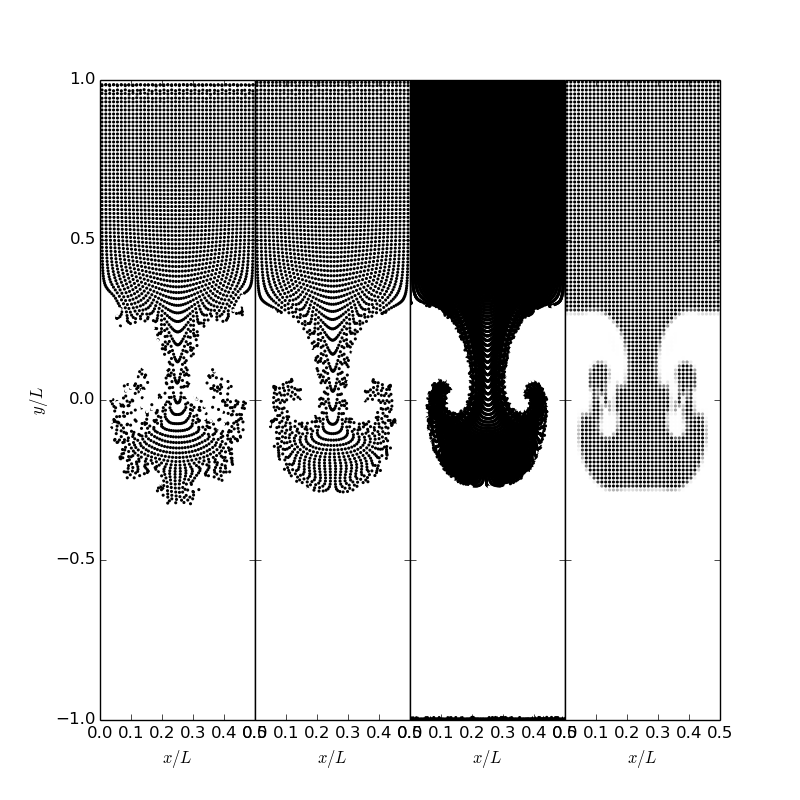}
  \caption{%
    Rayleigh-Taylor instability.  Color field at $t=1.32$, with black
    for particles with $\alpha=1$, and white (not visible, for clarity),
    for $\alpha=0$.  From left to right: partFEM method, projFEMq,
    projFEM6, and OpenFOAM (the latter in gray levels).  Only the
    middle section is shown, the other parts of the simulation being
    exactly repeated.
    \label{fig:RT_snaps}}
\end{figure}

In Fig. \ref{fig:RT_snaps} results are shown for the partFEM method, the
projFEMq, and the projFEM6.  Results from the OpenFOAM finite volume
method software are also included (see details in Appendix
\ref{sec:numerical}.)

The partFEM simulations seem to underestimate the viscosity, since
they give rise to an interface that is more distorted than the others.
The projFEMq and projFEM6 yield similar results, with the latter
producing smoother results at the cost of using $6$ particles per node
instead of just one.  These results may be compared against the
OpenFOAM result, which uses the same parameters and mesh resolution.
Note that OpenFOAM uses a volume-of-fluid algorithm which leads to
species diffusion, manifest in the appearance of gray nodes that have
intermediate values of the color field.  However, the $\alpha$ field,
when associated to the particles, should have in this problem crisp
values of either $0$ or $1$.  In addition, OpenFOAM results are
clearly affected by the underlying Eulerian mesh, as it is apparent
from the much less curved plume interface.  It therefore seems that
OpenFOAM results are affected by numerical artifacts in this case.
These artifacts are absent in the projection methods considered in
this work.

\section{Discussion}

There are two different points of view regarding the interpretation of
projPFEM methods.  In one of them, the simulation is Lagrangian in
spirit: the fluid particles move while carrying the fields, and the
mesh is merely used as a tool to evaluate space derivatives.  The
other approach is Eulerian in spirit: the mesh contains the fields and
the particles are only used to advect them from one time step to the
next one.  The resulting numerical algorithms are very similar if the
particles are kept from one time-step to the next, but the latter
point of view allows for greater flexibility in the treatment of the
particles: they may be re-created at (or, around) mesh nodes each time
steps, moved, and destroyed.  We have explored this possibility, and
the main difference is a better convergence (both as a function of
$\Delta t$ and of CPU time) of the velocity field at high resolutions,
while that of the pressure field is relatively unaffected.  Results
are not shown for the sake of brevity.  It is worth mentioning that
several other variants (such as ``projFEMq6'', with six times more
particles than mesh nodes, but quadratic functions for both, bottom
right results of Fig.  \ref{fig:zalesak}) have also been considered,
with results that seem better as far as convergence with $\Delta t$ is
concerned, but turn out to be quite similar when CPU time is
considered.

\section{Conclusions}
\label{sec:conclusion}
We have presented a combination of pFEM-like Lagrangian particles with
a standard FEM Eulerian mesh.  Our results confirm the findings of
Ref. \cite{Idelsohn_2015}, in which the projection errors, especially
from particles to mesh, are identified as the culprits of slow
convergence of the method. We restore an acceptable convergence rate
by including quadratically consistent shape functions for the
interpolation given simply in terms of products of standard linear
finite element basis functions.  The method is first validated on the
Zalesak's disk problem, which is used to screen possible variations on
the method.  These are later used on the Taylor-Green vortex sheet
flow, our main benchmark.  Convergence is not only tested as a
function of mesh refinement, but also in terms of CPU time, which is,
in our opinion, a critical and useful test to compare widely different
methods.

Our results show that the quadratic consistent method for projection
from mesh to particles and back gives a convergence superior to
previous projection methods in terms of the order of convergence and
in terms of CPU time. The method offers a very promising route to deal
with problems where a Lagrangian description is helpful (as in
fluid-structure interaction with large deformations, free surfaces,
sloshing, splashing, etc. \cite{PFEM,Idelsohn_2015}), and high
resolution results are required.

Of course, the procedure should be extended in many ways. Boundary
conditions, other than just periodic ones, should be considered. The
method should also be extended to 3D problems. Such an extension, even
if straightforward conceptually, could bring about technical
difficulties, specially in complicated geometries. In both 2D and 3D
application the method clearly relies on a rather uniform particle
distribution, a condition that may fail in some cases. A possible
remedy would be re-creating the particles periodically, as pointed out
in the previous section.

We finally stress once again that only a simple interpolation
procedure is used for projection. Other possibilities, such as
Galerkin projection, may provide superior results and are worth
considering in the future. The authors are also currently evaluating
the performance of procedures in which projection involves
convolutions with projection kernel functions.

\section*{Acknowledgements}
The research leading to these results has received funding from the
Ministerio de Econom\'{\i}a y Competitividad of Spain (MINECO) under
grants TRA2013-41096-P ``Optimizaci\'on del transporte de gas licuado
en buques LNG mediante estudios sobre interacci\'on
fluido-estructura'' and FIS2013-47350-C5-3-R ``Modelizaci\'on de la
Materia Blanda en M\'ultiples escalas''.

\bibliographystyle{wileyj}

\bibliography{sme_pdes,comp_geom}

\begin{thebibliography}{10}
\providecommand{\url}[1]{\texttt{#1}}
\providecommand{\urlprefix}{URL }
\expandafter\ifx\csname urlstyle\endcsname\relax
  \providecommand{\doi}[1]{doi:\discretionary{}{}{}#1}\else
  \providecommand{\doi}{doi:\discretionary{}{}{}\begingroup
  \urlstyle{rm}\Url}\fi

\bibitem{versteeg_2007}
Versteeg HK, Malalasekera W. \emph{An introduction to computational fluid
  dynamics: the finite volume method}. Pearson Education, 2007.

\bibitem{PIC}
Evans M, Harlow F. The particle-in-cell method for hydrodynamic calculations.
  \emph{Technical {R}eport LA-2139}, Los Alamos Scientific Laboratory 1957.

\bibitem{PIC2}
Amsden A. The particle-in-cell method for the calculation of the dynamics of
  compressible fluids. \emph{Technical {R}eport LA-3466}, Los Alamos Scientific
  Laboratory 1966.

\bibitem{chaniotis_2002}
Chaniotis AK, Poulikakos D, Koumoutsakos P. Remeshed smoothed particle
  hydrodynamics for the simulation of viscous and heat conducting flows.
  \emph{Journal of Computational Physics}  2002; \textbf{182}(1):67--90.

\bibitem{Feldman_Bonet_2007}
Feldman J, Bonet J. {Dynamic refinement and boundary contact forces in SPH with
  applications in fluid flow problems}. \emph{International Journal for
  Numerical Methods in Engineering}  2007; \textbf{72}(3):295--324.

\bibitem{marrone_2015}
Marrone S, Di~Mascio A, Le~Touz{\'e} D. Coupling of smoothed particle
  hydrodynamics with finite volume method for free-surface flows. \emph{Journal
  of Computational Physics}  2015; .

\bibitem{quinlan_2014}
Quinlan NJ, Lobovsk{\`y} L, Nestor RM. Development of the meshless finite
  volume particle method with exact and efficient calculation of interparticle
  area. \emph{Computer Physics Communications}  2014;
  \textbf{185}(6):1554--1563.

\bibitem{Idelsohn_2015}
Idelsohn S, O{\~n}ate E, Nigro N, Becker P, Gimenez J. {L}agrangian versus
  {E}ulerian integration errors. \emph{Computer Methods in Applied Mechanics
  and Engineering}  2015; \textbf{293}:191--206.

\bibitem{stanton2013non}
Stanton M, Sheng Y, Wicke M, Perazzi F, Yuen A, Narasimhan S, Treuille A.
  Non-polynomial {G}alerkin projection on deforming meshes. \emph{ACM
  Transactions on Graphics (TOG)}  2013; \textbf{32}(4):86.

\bibitem{Rapun2017}
Rap\'un ML, Terragni F, Vega JM. {LUPOD}: Collocation in {POD} via {LU}
  decomposition. \emph{Journal of Computational Physics}  2017; \textbf{335}:1
  -- 20.

\bibitem{PFEM}
O{\~n}ate E, Idelsohn SR, {del Pin} F, Aubry R. The particle finite element
  method - an overview. \emph{International Journal of Computational Methods}
  2004; \textbf{1}(2):267--307.

\bibitem{Liu_2003}
Liu M, Liu G, Lam K. Constructing smoothing functions in smoothed particle
  hydrodynamics with applications. \emph{Journal of Computational and Applied
  Mathematics}  2003; \textbf{155}(2):263 -- 284.

\bibitem{QCSF}
Duque D, Espa{\~n}ol P. Extending linear finite elements to quadratic precision
  in arbitrary meshes. \emph{Applied Mathematics and Computation}  2016;
  Accepted for publication.

\bibitem{codina_2001}
Codina R. Pressure stability in fractional step finite element methods for
  incompressible flows. \emph{Journal of Computational Physics}  2001;
  \textbf{170}(1):112--140.

\bibitem{CGAL}
{CGAL Editorial Board}. \textsc{Cgal}, {C}omputational {G}eometry {A}lgorithms
  {L}ibrary. Http://www.cgal.org.

\bibitem{Eigen}
Guennebaud G, Jacob B, \emph{et~al.}. Eigen v3. http://eigen.tuxfamily.org
  2010.

\bibitem{cholmod}
Chen Y, Davis TA, Hager WW, Rajamanickam S. Algorithm 887: Cholmod, supernodal
  sparse cholesky factorization and update/downdate. \emph{ACM Trans Math
  Software}  2009; \textbf{35}(3).

\bibitem{belikov1997}
Belikov V, Ivanov V, Kontorovich V, Korytnik S, Semenov AY. The non-{S}ibsonian
  interpolation: A new method of interpolation of the values of a function on
  an arbitrary set of points. \emph{Computational mathematics and mathematical
  physics}  1997; \textbf{37}(1):9--15.

\bibitem{arroyo2006}
Arroyo M, Ortiz M. Local maximum-entropy approximation schemes: a seamless
  bridge between finite elements and meshfree methods. \emph{International
  Journal for Numerical Methods in Engineering}  2006;
  \textbf{65}(13):2167--2202, \doi{10.1002/nme.1534}.
  \urlprefix\url{http://dx.doi.org/10.1002/nme.1534}.

\bibitem{cyron2009}
Cyron CJ, Arroyo M, Ortiz M. Smooth, second order, non-negative meshfree
  approximants selected by maximum entropy. \emph{International Journal for
  Numerical Methods in Engineering}  2009; \textbf{79}(13):1605--1632.

\bibitem{CGbook}
Okabe A, Boots B, Sugihara K, Chiu SN. \emph{Spatial tessellations: Concepts
  and applications of {V}oronoi diagrams}. 2nd edn., Probability and
  Statistics, Wiley: NYC, 2000. 671 pages.

\end{thebibliography}

\appendix

\section{Numerical methods}
\label{sec:numerical}

For all computational geometry procedures the CGAL 4.7 libraries
\cite{CGAL} are used.  In particular, the 2D Periodic Delaunay
Triangulation package, overloading the vertex base to contain the
relevant fields, and the face base to contain information relevant to
the edges.

The Eigen 3.0 linear algebra libraries \cite{Eigen} are also employed.
For the small linear algebra problem involved in the calculation of
the $A$ coefficients, SVD is used, with automatic rank detection. For
the large problems involved in the Galerkin procedure, the sparse
matrix package is used. The linear systems are solved iteratively for
partFEM, by the BiCGSTAB method.  For projFEM a direct method is
employed, with best results obtained using the CHOLMOD \cite{cholmod}
routines of the suitesparse project (through Eigen wrappers for
convenience, class CholmodSupernodalLLT). Slightly worse results are
obtained with eigen's build-in SimplicialLDLT class.

The OpenFOAM 3.0.1 finite volume method software has also been used.
Periodic (``cyclic'') boundary conditions are selected, on a 2D square
simulation cell. The utility funkySetFields, a part of the swak4Foam
extension package, is used to set the initial interface. In order to
mimic our simulations, we have modified the interFoam solver so that
only the pressure field is used (not the ``p\_rgh'' one, which would
include the gravity), and to add an additional term, $ (2 \alpha - 1)
\rho g$, to the right-hand side of the momentum equation.

Our computations took place on a 4-core Pentium 4 machine with 16 Gb
RAM.

\section{Quadratic correction} 
\label{sec:quad}

For an arbitrary arrangement of $N$ nodes labeled by the index $\mu$,
let us consider a set of functions $\{\phi_\mu\}$ that reconstruct
constant functions exactly (a property known as partition of unity):
\begin{equation}
  \label{eq:C0}
  \sum^N_{\mu=1}\phi_\mu(\bfr) = 1 .
\end{equation}
We will suppose that linear functions are also exactly reconstructed
by this set (a property known as the local co-ordinate):
\begin{equation}
  \label{eq:C1}
  \sum^N_{\mu=1} \phi_\mu(\bfr) \bfr_\mu = \bfr.
\end{equation}
These two properties combined define ``linear precision''.  These are
satisfied by many functional sets, including the well-known linear
finite elements (FE) shape functions, natural neighbor interpolants
(either Sibsonian or otherwise) \cite{belikov1997}, and LME and SME
interpolants \cite{arroyo2006,cyron2009}.

Only linear FE shape functions are considered here. In 1D, the nodal
functions for node $\nu$ is continuous and piece-wise linear, with
$\phi_\mu(x)$ going from $0$ at $x_{\mu-1}$ to $1$ at $x_\mu$, back to
$0$ at $x_{\mu+1}$. They are zero outside the segment
$(x_{\mu-1},x_{\mu+1})$. Similarly, in 2D $\phi_\mu(\bfr)$ is a
pyramid of height $1$ at node $\mu$ with straight edges that connect
the apex with the neighboring nodes. These neighbors are determined by
an underlying 2D triangulation; there are many possible ones, but we
choose, as is customary, the unique (except for possible degeneracies)
Delaunay triangulation \cite{CGbook}, thus defining natural neighbors.

The FE shape functions do \emph{not} comply with quadratic precision
in general:
\begin{equation}
  \label{eq:noC2}
  \sum^N_{\mu=1}\bfr_\mu \bfr_\mu^\mathrm{t} \phi_\mu(\bfr) \neq \bfr \bfr^\mathrm{t} .
\end{equation}

We here consider an ``extended'' set of shape functions derived from
the FE $\{\phi_{\mu}\}$ shape set, with functions:
\begin{align}
  \label{phiquad}
  \psi_{\mu}(\bfr) &:=\phi_\mu(\bfr)+
  \sum_{\nu\sigma}A_{\mu\nu\sigma}
  \phi_\nu(\bfr)\phi_\sigma(\bfr) ,
\end{align}
i.e. a linear combination of the original functions and their
pair-wise products. The inclusion of these products extends the
spatial support of the original functions (hence their name; the
number of functions in each set is the same.)

The coefficients will be taken to be symmetric with respect to the
last two indices: $A_{\mu\nu\sigma}=A_{\mu\sigma\nu}$, and are
determined so as to fulfill not only Eqs. (\ref{eq:C0}) and
(\ref{eq:C1}), but also to comply with quadratic precision:
\begin{equation}
  \label{eq:C2}
  \sum^N_{\mu=1}\bfr_\mu\bfr_\mu^\mathrm{t} \psi_\mu(\bfr) = \bfr\bfr^\mathrm{t} .
\end{equation}

One way to achieve this is by requiring that the coefficients
$A_{\mu\nu\sigma}$ satisfy
\begin{align}
  \label{eq:A0}
  \sum^N_{\mu=1} A_{\mu\nu\sigma}&=0
  \\
  \label{eq:A1}
  \sum^N_{\mu=1} \bfr_\mu A_{\mu\nu\sigma}&=0    \\
  \sum^N_{\mu=1} \bfr_\mu \bfr_\mu^\mathrm{t} A_{\mu\nu\sigma}&=
  \frac{1}{2}
  \left(
    \bfr_\nu \bfr_\sigma^\mathrm{t}+
    \bfr_\sigma \bfr_\nu^\mathrm{t} 
    -   \bfr_\nu\bfr_\nu^\mathrm{t}
    - \bfr_\sigma\bfr_\sigma^\mathrm{t}
  \right)
  \label{condquad}
\end{align}
as can be checked explicitly. 
The last identity can be rewritten with the help of the first two as
\begin{align}
  \label{eq:A2_2}
  \sum^N_{\mu=1} (\bfr_\mu-\bfr_\nu)(\bfr_\mu-\bfr_\sigma)^\mathrm{t} A_{\mu\nu\sigma}
  &=-\frac{1}{2}(\bfr_\nu-\bfr_\sigma)(\bfr_\nu-\bfr_\sigma)^\mathrm{t},
\end{align}
which is clearly translation and rotation invariant.

We want to have the largest number of coefficients $A_{\mu\nu\sigma}$
equal to zero and, if possible, only different from zero if
$\mu,\nu,\sigma$ are some kind of neighbors.  For every pair
$\nu,\sigma$ a set of equations must be fulfilled. It seems reasonable
to look only for pairs that are neighboring. Then, we require that
$A_{\mu\nu\sigma}=0$ if $\nu,\sigma$ are not neighbors or if
$\nu=\sigma$.  Then, we also require that $A_{\mu\nu\sigma}=0$ if
$\mu$ is not a neighbor (in some sense) of either $\nu$ or $\sigma$.

Notice that while this is, in principle, a purely geometrical
calculation that is not directly related to the original set of
functions, the particular definition of the geometrical concept of
``neighbor'' should be tied to the particular set employed. In the
case of standard FE examined in this work, this means Delaunay
neighbors (i.e. connected by an edge of the triangulation).  For a
full description of the procedure, we refer the reader to
\cite{QCSF}.

\end{document}